\documentclass[amssymb,aps,prb,twocolumn,superscriptaddress,showpacs]{revtex4}
\usepackage{graphicx}

\begin{document}
\title{Anisotropy of heavy hole spin splitting and interference effects of optical polarization
in semiconductor quantum wells subjected to an in-plane magnetic
field }
\author{Yuriy Semenov}
\affiliation{Institute of Semiconductor Physics,\\
National Academy of Sciences of Ukraine\\
Prospekt Nauki, 45 Kiev 03028 Ukraine}
\author{Sergij Ryabchenko}
\affiliation{Institute of Physics, National Academy of Sciences of Ukraine\\
Prospekt Nauki, 46 Kiev 03028 Ukraine}
\date{\today }

\begin{abstract}
Strong effects of optical polarization anisotropy observed
previously in the quantum wells subjected to the in-plane magnetic
field arrive at complete description within microscopic approach.
Theory we develop involves two sources of optical polarization.
First source is due to correlations between electron and heavy
hole (HH) phases of $\psi $-functions arising due to electron
Zeeman spin splitting and joint manifestation of low-symmetry and
Zeeman interactions of HH in an in-plane magnetic field. In this
case, four possible phase-controlled electron-HH transitions
constitute the polarization effect, which can reach its maximal
amount ( $\pm $1) at low temperatures when only one transition
survives. Other polarization source stems from the admixture of
excited light-holes (LH) states to HH by low-symmetry
interactions. The contribution of this mechanism to total
polarization is relatively small but it can be independent of
temperature and magnetic field. Analysis of different mechanisms
of HH splitting exhibits their strong polarization anisotropy.
Joint action of these mechanisms can result in new peculiarities,
which should be taken into account for explanation of different
experimental situations.
\end{abstract}


\maketitle

\section{Introduction}

The linear optical polarization $\rho $ of photo-luminescence (PL) in
quantum wells (QWs) are very sensitive to low symmetry interactions $V$,
which can be responsible for this polarization (Refs %
\cite{KusrPRL,Gaj,RyabchE,RyabchNATO}). A typical situation
corresponds to relatively weak $V$, which mixes the light hole
(LH) and heavy hole (HH) states. By this virtue the polarization
reaches the magnitude about $\varepsilon =\left| V\right| /\Delta
_{HL}$ ( $\Delta _{HL}$
is HH - LH energy splitting) without external magnetic field (Refs%
\cite{Gaj,RyabchE}).

Strong polarization of luminescence from [001] - oriented quantum wells Cd$%
_{1-x}$Mn$_{x}$Te /CdTe/ Cd$_{1-x}$Mn$_{x}$Te and its $\pi
$-periodic anisotropy (i.e. dependence on sample rotation about QW
normal) has been observed in Refs \cite{KusrPRL,RyabchNATO} under
in-plane magnetic field ${\vec{B}}$. There was assumed that these
properties are due to $C_{2v} $ symmetry potential of a hole in
QW. It was found that polarization and its anisotropy increases
sharply with increasing of in-plane magnetic field and reaches few
tens percents. This fact cannot be consistent with small value of
ratio $\varepsilon $. Moreover, strong polarization effects as
well as the significant contribution of fourth harmonic of
aforementioned anisotropy for narrow QWs remains so far
unexplained\cite{KusrPRL}. The phenomenological approach developed
in Ref\cite{KusrPRL} in terms of bilinear in ${\vec B}$
representation of $\rho $ cannot describe the strong effects when
$\rho \approx 1$.

The complementary approach of Ref.\cite{KusrPRL} in terms of
pseudospin formalism requires the proper determination of
pseudospin basic functions both for real electron spin operators
and for non-spin part of interaction being responsible for optical
transitions. Moreover different kinds of HH interactions, which
determine the HH splitting under in-plane magnetic field need
further consideration. This means that for correct description of
above experimental data it is necessary to have a microscopic
theory in terms of actual electron and hole spins rather then
pseudospins.

Here, we pay attention to the well-known fact that any interaction splitting
the degenerate electron and HH levels imposes some phase correlations
between the electron and hole wave functions. In addition to above small
contribution caused by LH-HH mixing, this correlation forms the polarization
and its anisotropy associated with some pair of distinguishable electron -
hole optical transitions regardless of spin levels splitting value.

On the other hand, there are different possible interactions which
are able to lift the HH degeneracy in a magnetic field
(Ref.\cite{Pikus}). Latter interactions impose their specific
correlations between electron and hole $\psi $ - function phases
as well as the period and phase of optical polarization anisotropy
(OPA), i.e. polarization dependence on QW rotation about its
normal. Thus, if one of the four possible electron - HH
transitions prevails (for instance, due to low enough
temperatures), one can expect the appearance of strong optical
polarization.

In this paper, we provide a quantitative microscopic analysis of
optical polarization anisotropy caused by different low-symmetry
interactions of a hole in QW. First, we discuss the general
expression for OPA in terms of electron and HH $\psi $- functions
phases. Then we show that the different interactions leading to HH
splitting reveal the various OPA dependencies on in-plane magnetic
field ${\vec B}$ rotation. This demonstrates the necessity to
account for joint contribution of aforementioned terms to OPA
resulting in qualitatively new peculiarities due to interference
effects. Finally, in the framework of our theory, we explain
quantitatively the most interesting experimental results that are
accessible from literature.

\section{Theoretical background}

\subsection{Photoluminescence linear polarization}

We are interested in the linear polarization of PL-spectrum that involves
four optical transitions from two electron spin sub-levels to two HH
sub-levels. To avoid the problems of these components spectral shifts in a
magnetic field (see below) we assume that integral PL intensities $I_{\alpha
}$ of polarization $\alpha $\ can be extracted from experiment and
associated with transition probabilities in terms of thermal population of
spin sub-levels. Thus, according to definition
\begin{equation}
\rho _{\alpha }=\frac{I_{\alpha }-I_{\alpha ^{\prime }}}{I_{\alpha
}+I_{\alpha ^{\prime }}},  \label{d1}
\end{equation}%
where the plane of $\alpha ^{\prime }$-polarization is perpendicular to that
of $\alpha $ polarization. Then, we introduce the reference frame associated
with main crystal axes so that ${\overrightarrow{OZ}}$ is parallel to growth
axis $[001]$, while $\overrightarrow{OX}\parallel \lbrack 100]$ and $%
\overrightarrow{OY}\parallel \lbrack 010]$ lay into QW plane.

The electron (or HH) spin splitting $\omega =\omega _{e}$ (or $\omega
=\omega _{h}$) is assumed to be described by following matrix Hamiltonian in
certain basis $\left\vert n\right\rangle $, $n=1$, $2$,
\begin{equation}
\left\Vert H_{n,n^{\prime }}\right\Vert =\frac{\omega }{2}\left(
\begin{array}{cc}
0 & e^{-i\theta } \\
e^{i\theta } & 0%
\end{array}%
\right) ,  \label{f1}
\end{equation}%
where $\omega =2\left\vert H_{1,2}\right\vert $, $\sin \theta =-2\ {\rm Im}%
H_{1,2}/\omega $. Eigenvalues and eigenfunctions of Hamiltonian (\ref{f1})
are
\begin{equation}
E_{\pm }=\pm \frac{1}{2}\omega ;\qquad \psi ^{\pm }=\frac{1}{\sqrt{2}}\left(
\pm e^{-i\theta /2}\left\vert 1\right\rangle +e^{i\theta /2}\left\vert
2\right\rangle \right) .  \label{f2}
\end{equation}

In the case of electron subjected to in-plane magnetic field ${\vec{B}}%
=B\{\cos \varphi ,\sin \varphi ,0\}$, $\psi ^{\pm }\equiv \psi _{c}^{\pm }$,
Hamiltonian $H=\omega _{e}{\vec{B}}{\vec{s}}$ takes the form (\ref{f1}) in
representation of $\left\vert 1\right\rangle =S\uparrow $ and $\left\vert
2\right\rangle =S\downarrow $, where $S$ is a periodic part of conductivity
band Bloch function, $\uparrow $ and $\downarrow $ are the eigenstates of
spin projection $s_{z}$. Here $\omega _{e}$ means{\it \ }electron Zeeman
spin splitting, $\theta =\varphi $\ is an angle between ${\vec{B}}$ and $%
\overrightarrow{OX}$ .

In the case of HH, $\psi ^{\pm }\equiv \psi _{v}^{\pm }$, the basis $%
\left\vert 1\right\rangle =L_{+}\uparrow $, $\left\vert 2\right\rangle
=-L_{-}\downarrow $ corresponds to $\pm 3/2$ projection of HH angular
momentum on $z$-direction, $L_{\pm }=\frac{1}{\sqrt{2}}(X\pm iY)$, $X$ and $%
Y $ are the periodic parts of valence band Bloch functions. The dependence $%
\theta =\theta (\varphi )$ have to be found for each specific form of HH
Hamiltonian (see below).

\begin{figure}[th]
\hspace*{20mm} 
\centering{\
\includegraphics[width=5cm]{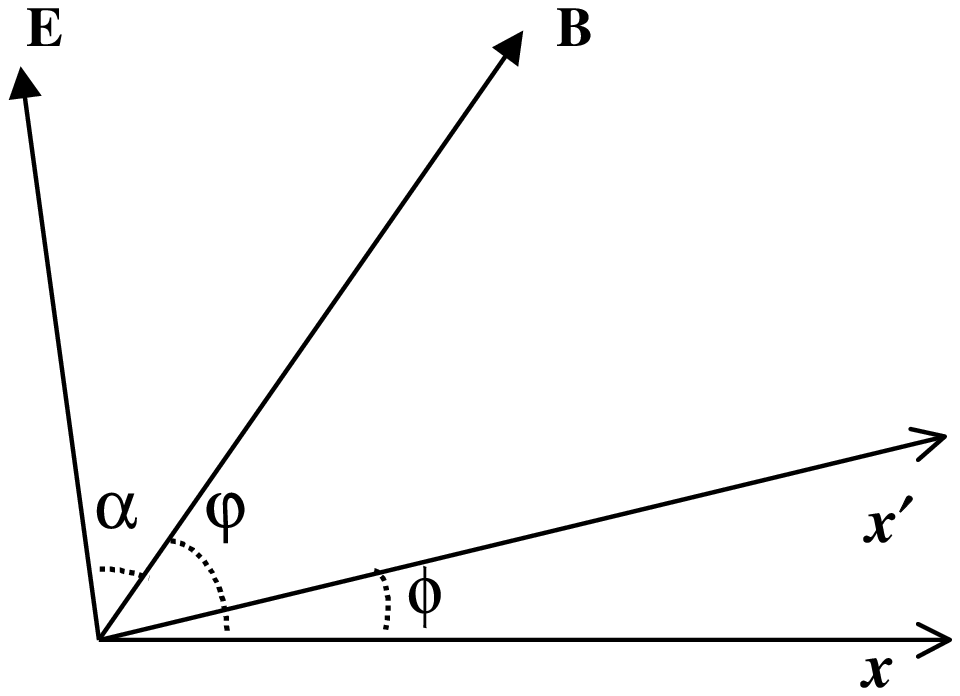}}
\caption{The relative positions of the crystal axis $x$, the axis
$x'$ of $C_{2v}$ interaction (\ref{f12}), direction of in-plane magnetic field ${%
\vec B}$ and the line $E$ of intersection of the plane of linear
polarization detection with the $(001)$ plane, and the angles that
between these lines.}
\end{figure}

The operator of interband optical transition with polarization plane rotated
relative to the magnetic field ${\vec{B}}=B\{\cos \varphi ,\sin \varphi ,0\}$
by angle $\alpha $ about the $\overrightarrow{OZ}$ axis (see Fig.1) takes
the form
\begin{equation}
\widehat{V}_{\alpha }=p_{-}e^{i(\varphi +\alpha )}+p_{+}e^{-i(\varphi
+\alpha )},  \label{f3}
\end{equation}%
where $p_{\pm }=\frac{1}{2}\left( e_{x}\pm ie_{y}\right) $, $e_{x}$ and $%
e_{y}$ are transformed as $x$ and $y$.

Using the definitions (\ref{f2}) and (\ref{f3}), one can easily find the
matrix element $M_{k,j}^{\alpha }=\left\langle \psi _{c}^{k}\left\vert
\widehat{V}_{\alpha }\right\vert \psi _{v}^{j}\right\rangle $ of
electro-dipole optical transition between electron states $\psi _{c}^{k}$, $%
k=\pm 1$, and HH states $\psi _{v}^{j}$, $j=\pm 1$, and corresponding
probability $W_{k,j}^{\alpha }=\left\vert M_{k,j}^{\alpha }\right\vert ^{2}$
\begin{equation}
W_{k,j}^{\alpha }\propto \left\{
\begin{array}{ll}
\sin ^{2}\left( 3\varphi /2+\alpha -\theta /2\right) , & k=j; \\
\cos ^{2}\left( 3\varphi /2+\alpha -\theta /2\right) , & k\neq j,%
\end{array}%
\right.  \label{f4}
\end{equation}%
where unimportant dimensional coefficient has been dropped. Similarly, one
can find the optical transition probability $W_{k,j}^{\alpha ^{\prime }}$
for perpendicular polarization plane that formally means the substitution $%
\alpha \rightarrow \alpha ^{\prime }=\alpha +\pi /2$ in Eq.(\ref{f4}). Since
the contribution of each of optical transitions $k\rightarrow j$ to total PL
intensity $I_{\alpha }$ is proportional to spin sub-level populations of
electron $P_{e}^{k}\propto e^{-k\omega _{e}/2T_{e}}/\left( e^{\omega
_{e}/2T_{e}}+e^{-\omega _{e}/2T_{e}}\right) $ and HH $P_{h}^{j}\propto
e^{-j\omega _{h}/2T_{h}}/\left( e^{\omega _{h}/2T_{h}}+e^{-\omega
_{h}/2T_{h}}\right) $ ($T_{e}$ and $T_{h}$ are the electron and HH spin
temperatures in energy units that can differ from lattice temperature $T$),
general Eq. (\ref{d1}) takes following form
\begin{equation}
\rho _{\alpha }^{(0)}=\frac{%
\mathop{\displaystyle\sum}%
\limits_{k,j}P_{e}^{k}P_{h}^{j}\left( W_{k,j}^{\alpha }-W_{k,j}^{\alpha
^{\prime }}\right) }{%
\mathop{\displaystyle\sum}%
\limits_{k,j}P_{e}^{k}P_{h}^{j}\left( W_{k,j}^{\alpha }+W_{k,j}^{\alpha
^{\prime }}\right) }.  \label{f5}
\end{equation}

Substitution of Eq. (\ref{f4}) and expressions for $P_{e}^{k}$ and $%
P_{h}^{j} $ into Eq. (\ref{f5}) leads after some algebra to following simple
result
\begin{eqnarray}
\rho _{\alpha }^{(0)} &=&-P_{eh}\cos \left( 3\varphi +2\alpha -\theta
\right) ;  \label{f6a} \\
P_{eh} &=&\tanh \left( \omega _{e}/2T_{e}\right) \tanh \left( \omega
_{h}/2T_{h}\right) .  \label{f6c}
\end{eqnarray}

Notice that Eq. (\ref{f6a}) does not describe all possible polarization
effects in QW. A closer look at the derivation of Eq. (\ref{f6a}) shows that
low symmetry perturbations $V$ of HH basis wave functions $\left\vert
1\right\rangle $ and $\left\vert 2\right\rangle $ should also be taken into
account along with HH splitting in spite of small value $\varepsilon
=\left\vert \left\langle m\left\vert V\right\vert m\pm \Delta m\right\rangle
\right\vert /\Delta _{HL}$, where $\left\vert m\right\rangle $ and $%
\left\vert m\mp \Delta m\right\rangle $ are non perturbed HH ($\left\vert
m\right\vert =3/2$) and LH ($\left\vert m\mp \Delta m\right\vert =1/2$)
basis functions. Doing so needs a distinction between the case $\Delta m=1$,
leading to HH splitting in the third order and the case $\Delta m=2$,
leading to formation of effective $g$- factor in the first order. Thus, the
perturbation $V$ gives rise to corrections $\delta \rho _{\alpha }\sim
\varepsilon ^{3-\Delta m}$ to total polarization that can be now written as
a sum
\begin{equation}
\rho _{\alpha }=\rho _{\alpha }^{(0)}+\delta \rho _{\alpha }.  \label{f7}
\end{equation}

Explicit form of $\delta \rho _{\alpha }$ depends on specific form
of interaction leading to HH-LH mixing.

Comparing two contributions of different mechanisms to Eq. (\ref{f7}), let
us note that electron-HH spin correlations ($\rho _{\alpha }^{(0)}$-term)
dominate in OPA at sufficiently low temperatures. However, the $\delta \rho
_{\alpha }$ -term can dominate at high temperatures or zero (small) magnetic
field. In the following, we concentrate on two important cases: polarization
$\rho _{0}$ along a magnetic field direction with $\alpha =0^{\circ }$ and
polarization $\rho _{45}$ in the plane rotated relatively $\overrightarrow{B}
$ by $\alpha =45^{\circ }$.

\subsection{Spectral dependence of linear polarization}

In this subsection we discuss the effects of spectral shifts of electron-HH
optical transitions caused by spin splitting (\ref{f2}){\it .} A simplest
situation corresponds to the electron-HH optical line splitting into the
four plainly distinguishable components with polarizations $\rho _{\alpha
k,j}=\left( W_{k,j}^{\alpha }-W_{k,j}^{\alpha ^{\prime }}\right) /\left(
W_{k,j}^{\alpha }+W_{k,j}^{\alpha ^{\prime }}\right) =-kj\cos \left(
3\varphi +2\alpha -\theta \right) $ and intensities $I_{kj}\propto
P_{e}^{k}P_{h}^{j}$. If these components overlap with each other (i.e.
splitting of spectral components are smaller than their linewidth $\sigma
_{kj}$ ), the polarization depends on spectral position at the contour of a
composite line of optical transitions. Thus, it is convenient to determine
the spectral-dependent polarization
\begin{equation}
\rho _{\alpha }\left( \omega \right) =\frac{I_{\alpha }\left( \omega \right)
-I_{\alpha ^{\prime }}\left( \omega \right) }{I_{\alpha }\left( \omega
\right) +I_{\alpha ^{\prime }}\left( \omega \right) }.  \label{s1}
\end{equation}%
The intensity $I_{\alpha }\left( \omega \right) $ of optical transition at
the frequency $\omega $ depends on lineshape of each electron - HH
transition $f_{k,j}\left( \omega \right) =f(\omega -(\omega _{0}+k\frac{%
\omega _{e}}{2}-j\frac{\omega _{h}}{2}))$, where possible dependence of $%
\omega _{0}$ on magnetic field describes the effect of the lines
center-of-mass shift in a magnetic field ${\vec{B}}$. We assume also that
linshapes $f_{k,j}\left( \omega \right) $ with linewidth $\sigma
_{kj}=\sigma $\ are same for all transitions $\left\vert k\right\rangle
\rightarrow \left\vert j\right\rangle $. So, Eq. (\ref{s1}) takes the form
\begin{equation}
\rho _{\alpha }^{(0)}\left( \omega \right) =\frac{\sum\limits_{k,j}f_{k,j}%
\left( \omega \right) \left( W_{k,j}^{\alpha }-W_{k,j}^{\alpha ^{\prime
}}\right) P_{e}^{k}P_{h}^{j}}{\sum\limits_{k,j}f_{k,j}\left( \omega \right)
\left( W_{k,j}^{\alpha }+W_{k,j}^{\alpha ^{\prime }}\right)
P_{e}^{k}P_{h}^{j}}.  \label{s2}
\end{equation}

In the case of small magnetic field shifts $\omega _{e}$ and $\omega _{h}\ll
\sigma $, the lineshape function can be expanded into power series
\begin{widetext}
\begin{equation}
f_{k,j}\left( \omega \right) \approx f(\Delta \omega )-\frac{1}{2}f^{\prime
}(\Delta \omega )(k\omega _{e}-j\omega _{h})+\frac{1}{8}f^{\prime \prime
}(\Delta \omega )(k\omega _{e}-j\omega _{h})^{2},  \label{s3}
\end{equation}
\end{widetext}
where $f'(\omega )$ and $f''(\omega )$ are the first
and second derivatives of $f(\omega )$, $\Delta \omega =\omega -\omega _{0}$%
. Substitution of this expansion into the Eq. (\ref{s2}) with regard to Eq. (%
\ref{f4}) results in net effect similar to that of Eq. (\ref{f6a})\ where $%
P_{eh}$ should be changed by
\begin{widetext}
\begin{eqnarray}
&&P_{T}(\omega ) =\tanh \left( \frac{\omega _{e}}{2T_{e}}\right)
\tanh
\left( \frac{\omega _{h}}{2T_{h}}\right) -  \nonumber \\
&&- \frac{\sigma f^{\prime }(\Delta \omega )}{2f(\Delta \omega
)}\left[
\frac{\omega _{h}}{\sigma }\tanh \left( \frac{\omega _{e}}{2T_{e}}\right) -%
\frac{\omega _{e}}{\sigma }\tanh \left( \frac{\omega _{h}}{2T_{h}}\right) %
\right] -\frac{\sigma ^{2}f^{\prime \prime }(\Delta \omega )}{4f(\Delta
\omega )}\frac{\omega _{e}\omega _{h}}{\sigma ^{2}}.  \label{s4}
\end{eqnarray}
\end{widetext}
Latter equation displays sharp polarization dependence on the detuning $%
\Delta \omega $. Moreover, this dependence is determined by
specific lineshape. In the case of Gaussian or Lorentzian shape
$P_{T}(\omega
)\simeq P_{eh}+\left( \frac{1}{2}-\frac{\Delta \omega ^{2}}{\sigma ^{2}}%
\right) \frac{\omega _{e}\omega _{h}}{\sigma ^{2}}$ or $P_{T}(\omega )\simeq
P_{eh}+\frac{\sigma ^{2}-3\Delta \omega ^{2}}{2(\sigma ^{2}+\Delta \omega
^{2})^{2}}\frac{\omega _{e}\omega _{h}}{\sigma ^{2}}$, respectively.{\small %
\ }

Finally, we have to consider the intermediate case, $\omega _{h}\ll \sigma
\ll \omega _{e}$. This is because inequality $\omega _{h}\ll \omega _{e}$
usually takes place in a wide range of the magnetic fields. Two components $%
k=\pm $ of different electron spin states have\ intensities $I_{k}\propto
P_{e}^{k}\ $and opposite signs of polarization with OPA in the form similar
to (\ref{f6a}), with
\begin{equation}
P_{T,k}(\omega )=k(\tanh \left( \omega _{h}/2T_{h}\right) -\frac{\sigma
f^{\prime }(\Delta \omega )}{2f(\Delta \omega )}\frac{\omega _{h}}{\sigma }).
\label{s5}
\end{equation}%
to be substituted instead $P_{eh}$.

The difference between Gaussian and Lorentzian shapes of spectral lines
displays, respectively, $P_{T,k}(\omega )\simeq k(\tanh \left( \omega
_{h}/2T_{h}\right) +\Delta \omega \omega _{h}/\sigma ^{2})$ and $%
P_{T,k}(\omega )\simeq k(\tanh \left( \omega _{h}/2T_{h}\right) +\Delta
\omega \omega _{h}/(\Delta \omega ^{2}+\sigma ^{2}))$ that is evident at the
wing of the lines.

Evidently, an in-plane magnetic field contributes to linear polarization of
the optical spectra of absorption, reflectivity, etc. In these cases, the
mechanism of wave functions phase correlations can become apparent due to HH
and electron spin splitting also. As distinct from PL, in these cases
optical transitions occur between completely populated valence electron
states and empty conductivity electron ones. Therefore, Eqs (\ref{s4}) and (%
\ref{s5}) describe this situation in the limit $T_{e}$\ and $%
T_{h}\rightarrow \infty $. However the polarization of reflectivity spectra
should be described in terms of standard equation for reflection coefficient
with transition probabilities (\ref{f4}).

In subsequent calculations, we primarily focus on PL polarization (Eqs (\ref%
{f6a}) and (\ref{f6c})) since it has been thoroughly studied experimentally
in the literature.

\section{The HH interactions}

We consider sequentially the HH interactions according to lowering the
symmetry of QW potential. In doing so these interactions assume to be small
perturbations with respect to $\Delta _{HL}$.

\subsection{Zeeman interaction}

Zeeman interaction in terms of hole effective angular momentum $J=3/2$ is
isotropic
\begin{equation}
V_{Z}={\vec{G}}_{{h}}{\vec{J}}=G_{h}(J_{X}\cos \varphi +J_{Y}\sin \varphi ).
\label{f8}
\end{equation}

Here ${\vec{G}}_{{h}}$ is an effective\ in-plane magnetic field in
energy units that can include the effects of carrier-ion
(hole-ion) exchange interaction in the case of DMS quantum
structure.\cite{DMSQW,DMS2} In the case of $[001]$-orientated QW
$V_{Z}$ does not split the HH states in first and second orders in
perturbation. Third order can be represented by effective
Hamiltonian with Pauli matrices ${\vec{\sigma}}$ in terms of basis
functions $\left\vert 1\right\rangle $ and $\left\vert
2\right\rangle $, calculated in second order of perturbation
theory according to L\"{o}wdin procedure (see Ref.\cite{BirPikus})
\begin{equation}
V_{Z}^{(3)}=\frac{3}{4}\Delta _{HL}h^{3}(\sigma _{x}\cos 3\varphi +\sigma
_{y}\sin 3\varphi ),  \label{f9}
\end{equation}%
where we introduce dimensionless parameter $h=G_{h}/\Delta _{HL}$.

The Eq. (\ref{f9}) gives {\em isotropic} HH splitting $\omega _{Z}=\frac{3}{2%
}\Delta _{HL}h^{3}$ and $\psi $-function phase $\theta =3\varphi $.
According to Eq. (\ref{f6a}) this corresponds to isotropic polarization $%
\rho _{\alpha }^{(0)}=-P_{eh}\cos 2\alpha $. The polarization is maximal
along or across magnetic field direction ($\alpha =0$ or $90^{\circ }$). The
polarization is absent for $\alpha =45^{\circ }$, that can be also expected
from symmetry considerations. If magnetic field is weak enough, one can find
$P_{eh}\simeq G_{h}^{3}G_{e}/T_{e}T_{h}\Delta _{HL}^{2}\propto B^{4}$. In
this case the contribution from LH admixture can be more important. The
calculation of LH contribution to HH polarization stemming from LH admixture
to the basis functions $\left\vert 1\right\rangle $ and $\left\vert
2\right\rangle $ gives rise to the corrections $\delta \rho
_{0}^{(2)}=-h^{2} $ and $\delta \rho _{45}^{(2)}=0$.

\subsection{Non Zeeman interaction with a magnetic field}

Symmetry of Luttinger Hamiltonian admits the existence of non-Zeeman
interaction of holes with a magnetic field in the form
\begin{equation}
V_{q}=q_{1}G_{h}\left( J_{x}^{3}\cos \varphi +J_{y}^{3}\sin \varphi \right) ,
\label{f10}
\end{equation}%
where $q_{1}$ is a relatively small parameter reflecting the interaction
between valence and $\Gamma _{15}$ -conductivity bands (Ref.\cite{Ivch}%
). This Hamiltonian has non-zero matrix elements between HH states $%
\left\vert 3/2\right\rangle $ and $\left\vert -3/2\right\rangle $ that
defines the effective HH Hamiltonian in first order in perturbation (\ref%
{f10})
\begin{equation}
V_{q}^{(1)}=\frac{3}{4}\Delta _{HL}q_{1}h\left( \sigma _{x}\cos \varphi
-\sigma _{y}\sin \varphi \right) .  \label{f11}
\end{equation}%
Comparison of Eq. (\ref{f11}) with Eq. (\ref{f1}) gives {\em isotropic} HH
splitting $\omega _{q}=\frac{3}{2}q_{1}G_{h}$ and $\psi $-function phase $%
\theta =-\varphi $. Therefore interaction (\ref{f10}) can be responsible for
fourth harmonic of OPA (\ref{f6a}), $\rho _{\alpha }^{(0)}=-P_{eh}\cos
(4\varphi +2\alpha )$, which correlate with a cubic anisotropy of Luttinger
Hamiltonian. Note that in-plane g-tensor $g_{\mu \nu }^{\perp }$ that can be
defined in terms of Eq. (\ref{f11}) for HH pseudospin\cite{Ivch} $\widetilde{%
s}_{x}=\sigma _{x}/2$ and $\widetilde{s}_{y}=-\sigma _{y}/2$ is isotropic
i.e. $g_{xx}^{\perp }=g_{yy}^{\perp }$, $g_{xy}^{\perp }=g_{yx}^{\perp }=0$.

\subsection{Potentials of C$_{2v}$ symmetry}

Most OPA experiments performed up to now have found some $\pi
$-periodic component of OPA. It was assumed, that such kind of
anisotropy is due to the hole potentials of $C_{2v}$ symmetry.
\cite{KusrPRL,Gaj,RyabchNATO,Pikus} We consider two reasons for
appearance of $C_{2v}$ hole potential in QW. First is a $C_{2v}$
constituent (so-called interface $C_{2v}$ potential) of
heterojunction potential inherent in [001] oriented structures
composed of zinc-blend
semiconductors.\cite{IvchVoisin,Voisin,YakovIvch} In QW structures
with common anion (cation), the contributions of two interface
potentials compensate each other. However, this compensation is
not complete in the case of non-identical barriers or interface
profile.\cite{Gaj} This kind of
interaction can be written in terms of hole angular moment in the form \cite%
{IvchVoisin} $V_{if}=t_{if}\{J_{x},J_{y}\}$, where $t_{if}$ is an
interaction constant ($\left\vert t_{if}\right\vert \ll \Delta _{HL}$), $%
\{J_{x},J_{y}\}=(J_{x}J_{y}+J_{y}J_{x})/2$. Additionally to $V_{if}$, there
is also $C_{2v}$ potential $V_{d}=d\varepsilon
_{x_{d}y_{d}}\{J_{x_{d}},J_{y_{d}}\}$ caused by in-plane strains.\cite{Pikus}
Here $d$ is a deformation potential; $\varepsilon _{x_{d}y_{d}}$ is the
strain with $x_{d}$ and $y_{d}$ principal axes forming some angle with $[100]
$ and $[010]$ directions. Actually, we do not need to consider $V_{if}$ and $%
V_{d}$ separately since their sum is also $C_{2v}$ potential $V_{t}$, which
takes the canonical form in terms of total amplitude $T_{t}$ and axes $%
x^{\prime }$ and $y^{\prime }$ forming angle $\phi $ with $[100]$ and $[010]$
directions
\begin{equation}
V_{t}=T_{t}\{J_{x'},J_{y'}\}.  \label{f12}
\end{equation}%
As an illustration, Fig.1 shows a position of the coordinate axes defining
the angles $\varphi $, $\phi $ and $\alpha $.

The potential (\ref{f12}) does not lift the $\pm 3/2$ HH degeneracy but
results in $\mp 1/2$ LH admixture in first order of perturbation theory.
This generates some temperature and magnetic field independent polarization
\cite{Gaj} $\delta \rho _{\alpha }^{(1)}$ with respect to polarization plane
$\alpha $: $\delta \rho _{\alpha }^{(1)}=-t\sin 2(\varphi -\phi +\alpha )$,
where $t=T_{t}/\Delta _{HL}$.\cite{comm2}

In the presence of a magnetic field, the potential $V_{t}$ generates an
effective in-plane $g$-factor for HHs.\cite{Pikus} This effect can be taken
into account in lowest order as interference of $V_{Z}$ (\ref{f8}) and $%
V_{t} $ (\ref{f12}). In terms of Pauli matrices, the HH splitting is
described by the effective Hamiltonian
\begin{equation}
V_{ht}^{(2)}=-\frac{3}{2}\Delta _{HL}ht\left( \sigma _{x}\sin \left( \varphi
+2\phi \right) -\sigma _{y}\cos \left( \varphi +2\phi \right) \right) ,
\label{f13}
\end{equation}%
which defines the phase $\theta =\varphi +2\phi +\pi /2$ and {\em isotropic}
HH splitting $\omega _{ht}=3\Delta _{HL}ht$. In spite of this fact,
representation of Eq.(\ref{f13}) in the form of Zeeman interaction for
pseudospin $\widetilde{s}$ determines anisotropic g-tensor. In terms of $%
x^{\prime }$ and $y^{\prime }$ reference system, one can find
$g_{x^{\prime }x^{\prime }}^{\perp }=g_{y'y'}^{\perp }=0$,
$g_{x^{\prime }y^{\prime }}^{\perp }=g_{y^{\prime }x^{\prime
}}^{\perp }$.\cite{comm3} So, the effect of $C_{2v}$ -OPA
(\ref{f6a}) is described by $\rho _{\alpha }^{(0)}=-P_{eh}\sin
2\left( \varphi -\phi +\alpha \right) $. If a magnetic field is
sufficiently weak, one can find $P_{eh}\simeq
3G_{h}G_{e}/4T_{e}T_{h}\propto B^{2}$.\cite{KusrPRL} Note once
again that the nature of polarization $\rho _{\alpha }^{(0)}$
calculated with non-perturbed basic functions $\left\vert \pm
3/2\right\rangle $ is different from that of $\delta \rho _{\alpha
}^{(1)}$ calculated with LH
admixture in first order. Thus, one can easily imagine a situation when $%
\rho _{\alpha }^{(0)}\gg \delta \rho _{\alpha }^{(1)}$ despite the fact that
$V_{ht}^{(2)}$ generates $\rho _{\alpha }^{(0)}$ in second perturbation
order in $V_{Z}$ and $V_{t}$, while $\delta \rho _{\alpha }^{(1)}$ arises in
first order in $V_{t}$.

\subsection{Random potential of HH localization}

The PL in QWs is known to come from the localized electronic (hole) states,
which are formed by a random potential (interface roughness, defects, etc).
In general case, the profile of this potential (and therefore the hole
density) is not symmetric in QW plane. In-plane asymmetry of localized hole $%
\psi $-function leads to mixing of HH $\Psi _{H}$ and LH $\Psi _{L}$ states
that results in appearance of a finite HH $g$-factor (Ref.\cite{Ivch}%
). Corresponding Hamiltonian, describing the splitting of
localized HHs can be represented as some $C_{2v}$ potential (see
Ref.\cite{Sem02} for details)
\begin{equation}
V_{h\varkappa }^{(2)}=\Delta _{HL}h\varkappa \left( \sigma _{x}\cos \left(
\varphi +2\phi _{\varkappa }\right) +\sigma _{y}\sin \left( \varphi +2\phi
_{\varkappa }\right) \right) .  \label{r1}
\end{equation}%
Here $\phi _{\varkappa }$ determines the axes $\xi $ and $\eta $ for
canonical form (\ref{f12}) representation of $V_{h\varkappa }^{(2)}$
potential by means of an angle between $x^{\prime }$ and $\xi $, $\varkappa =%
\overline{\gamma }\hbar ^{2}(K_{\xi }^{2}-K_{\eta }^{2})/m_{0}\Delta _{HL}$
and $K_{\alpha }^{2}=\left\langle \Psi _{H}\mid \Psi _{L}\right\rangle
\left\langle \Psi _{L}\left\vert -\partial ^{2}/\partial \nu ^{2}\right\vert
\Psi _{H}\right\rangle $, $\nu =\xi $, $\eta $; $\gamma _{2}<\overline{%
\gamma }<\gamma _{3}$, $\gamma _{2}$ and $\gamma _{3}$ are Luttinger
parameters. In the case of axially symmetric $\psi $ function, the equality $%
K_{\xi }^{2}=K_{\eta }^{2}$ gives zero $V_{h\varkappa }^{(2)}$ potential. In
general case, $K_{\xi }^{2}\neq K_{\eta }^{2}$ , the potentials (\ref{f13})
and (\ref{r1}) can be combined into single quantity $V_{ht\varkappa }^{(2)}$
that takes the form (\ref{f13}) with effective amplitude $t_{\varkappa }$
and angle $\phi _{r}$ instead of $t$ and $\phi $:
\begin{eqnarray}
t_{\varkappa } &=&\sqrt{\left( t+\varkappa \cos 2\phi _{\varkappa }\right)
^{2}+\left( \varkappa \sin 2\phi _{\varkappa }\right) ^{2}},  \label{r2} \\
\phi _{r} &=&\frac{1}{2}\arcsin \frac{\varkappa \sin 2\phi _{\varkappa }}{%
t_{\varkappa }}+\phi .  \label{r3}
\end{eqnarray}%
It should be emphasized that the Eq. (\ref{r1}) has to do with a
single hole localized on some fluctuation of random QW potential.
The observable PL polarization is the result of addition of a
great number of localized state contributions with random
amplitudes and principal axes directions. Thus, the complete
description of PL polarization has to include an averaging over
the parameters $\varkappa $ and $\phi _{\varkappa }$ with some
distribution functions. If there is no preferential directions for
HH localization in QW plane, the potential $V_{h\varkappa }^{(2)}$
cannot lead to OPA. Moreover, numerical analysis shows that random
potential (\ref{r1}) can greatly suppress the magnitude of
polarization and therefore amplitude of OPA as soon as it exceeds
other regular (non-random) interactions. In subsequent discussion
we omit the contribution of $V_{h\varkappa }^{(2)}$ into OPA for
simplicity. Nevertheless, taking into account the random potential
influence may be in need for a description of realistic
experimental situations.

Magnetic polaron effect\cite{Yakovlev} should also be considered as an
intensification factor for the parameter $\varkappa $.

\section{Interference effects}

Foregoing analysis shows few important polarization properties attributed to
different kinds of HH interactions in QW. We have found that HH splitting is
isotropic for each HH Hamiltonian (\ref{f9}), (\ref{f11}) and (\ref{f13}).
However, they reveal different OPA's that points to possible interplay among
these contributions. Let us consider joint manifestation of interactions
that can split HH states
\begin{equation}
V=V_{Z}^{(3)}+V_{q}^{(1)}+V_{ht}^{(2)}.  \label{f14}
\end{equation}%
First, we should find the total module $\omega _{h}/2$ and phase $\theta $
for matrix element
\begin{equation}
V_{1,2}=\frac{1}{2}\left( \omega _{Z}e^{-i3\varphi }+\omega _{q}e^{i\varphi
}+\omega _{ht}e^{-i(\varphi +2\phi +\pi /2)}\right) .  \label{f15}
\end{equation}%
After some algebra, the HH splitting can be written in terms of two
expressions $Q$ and $R$%
\begin{eqnarray}
\omega _{h} &=&\frac{3}{2}\Delta _{HL}h\sqrt{Q^{2}+R^{2}};  \label{f16} \\
Q &=&2t\cos 2\left( \varphi -\phi \right) -q_{1}\sin 4\varphi ;  \label{f16a}
\\
R &=&h^{2}+2t\sin 2\left( \varphi -\phi \right) +q_{1}\cos 4\varphi .
\label{f16b}
\end{eqnarray}%
In a similar manner, we calculate the polarization (\ref{f6a}) where $\theta
$ have to be found from definition (\ref{f15}). Thus, after some identical
manipulations, we obtain $\rho _{\alpha }$ for two values of $\alpha $
\begin{equation}
\rho _{\alpha }=\left\{
\begin{array}{ll}
-P_{eh}\frac{R}{\sqrt{Q^{2}+R^{2}}}-h^{2}-t\sin 2(\varphi -\phi ), & \alpha
=0^{\circ }; \\
-P_{eh}\frac{Q}{\sqrt{Q^{2}+R^{2}}}-t\cos 2(\varphi -\phi ), & \alpha
=45^{\circ }.%
\end{array}%
\right.  \label{f17}
\end{equation}%
Here for completeness, we take also into account the corrections for LH-HH
mixing. The Eq.(\ref{f17}) with Eqs (\ref{f6c}) and (\ref{f16}) are the
final results of our calculations that cover the most practically important
cases.

One can see that HH splitting (\ref{f16}) reveals magnetic anisotropy with
finite magnitudes of $t$ and $q_{1}$ despite the isotropic character of HH
splitting of each of terms (\ref{f14}) taken separately. Moreover, the
effective HH transversal $g$-factor $\widetilde{g}_{\perp }=\omega _{h}/G_{h}
$ can be turned to zero at some direction and amplitude of a magnetic field.
To demonstrate this fact let us note that equation $\widetilde{g}_{\perp }=%
\frac{3}{2}\sqrt{Q^{2}+R^{2}}=0$ agrees with the system of two equations $Q=0
$ and $R=0$ in variables $\varphi $ (the angle) and $h$ (amplitude).
Simplest solution of this system can be obtained in the case of $\phi =0$,
namely $\varphi =\pm \pi /4$ and $h=\sqrt{\pm 2t+q_{1}}$.

\section{Discussion. Comparison with experiment}

As it is obvious from the Eq.(\ref{f17}), the anisotropy and possible random
degeneration of HH splitting do not influence OPA for high temperatures $%
T_{h}\gg \omega _{h}$, (or $P_{eh}\propto \omega _{h}$). In this case we may
expect the additive contributions of different aforementioned OPA
mechanisms: $\rho _{0}^{(0)}\simeq -\frac{3}{2}R\frac{G_{h}}{T_{h}}\tanh
\left( \omega _{e}/2T_{e}\right) $, $\rho _{45}^{(0)}\simeq -\frac{3}{2}Q%
\frac{G_{h}}{T_{h}}\tanh \left( \omega _{e}/2T_{e}\right) $, i.e. the
amplitude of second harmonic of OPA is proportional to $t$ while that of
fourth harmonic is proportional to $q_{1}$.
\begin{figure}[th]
\centering{\
\includegraphics[width=6cm]{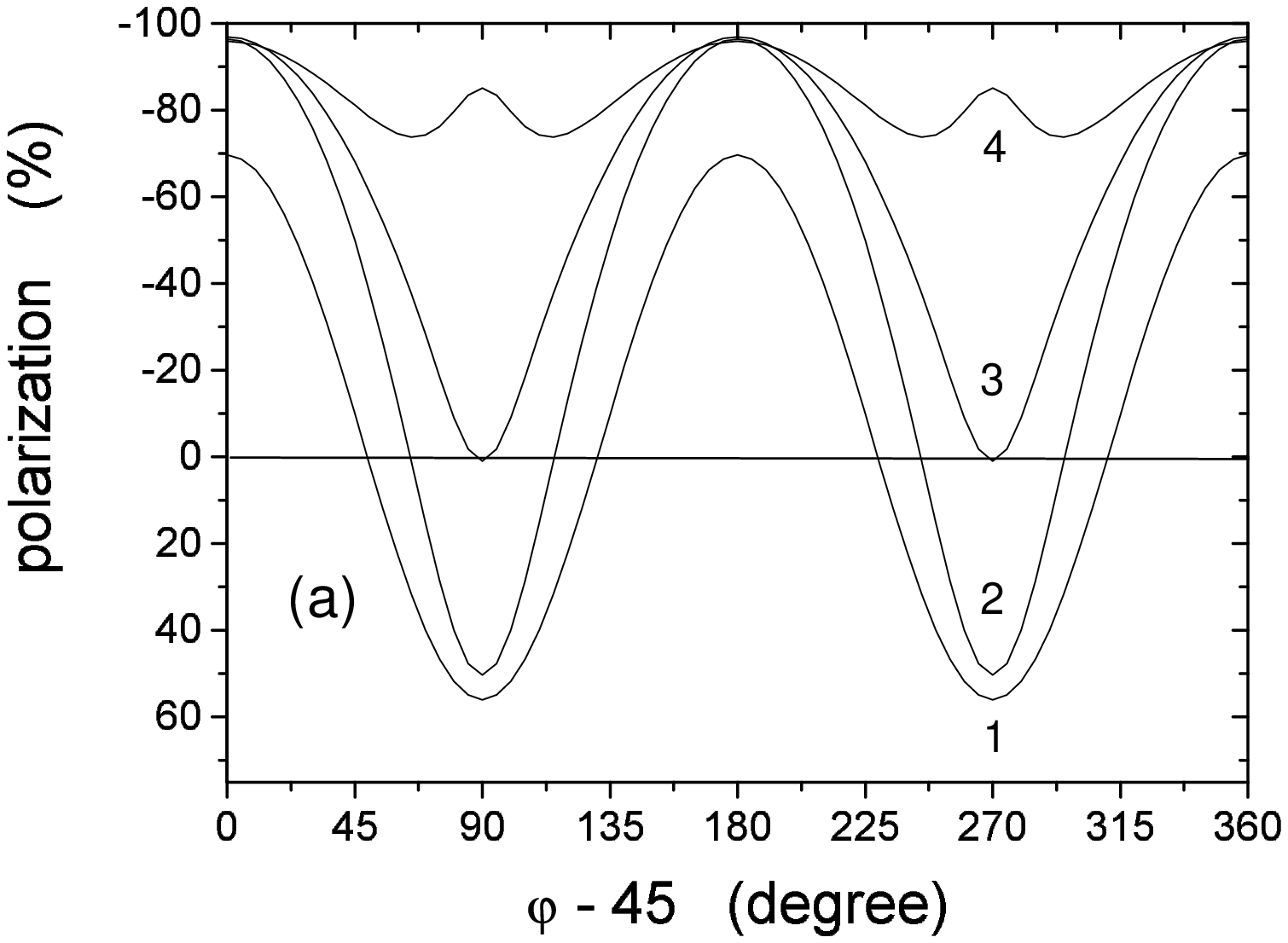}}
\vspace*{15mm}
\centering{\
\includegraphics[width=6cm]{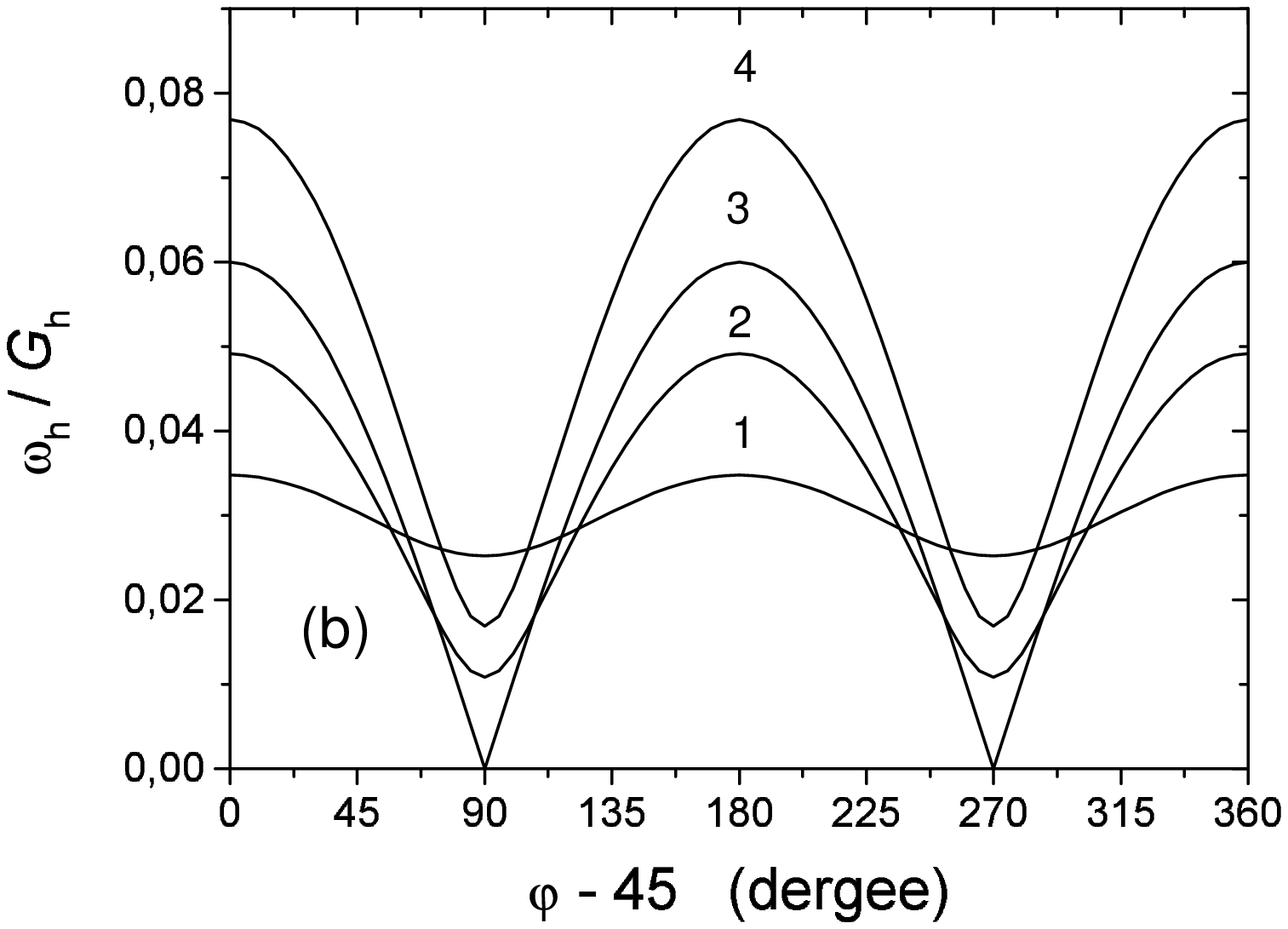}}
\caption{The OPA (a) and transversal effective HH $g$-factor $\widetilde{g}%
_{\perp }=\omega _{h}/G_{h}$ (b) calculated for $q_{1}=0$,
$t=0.01$, $\Delta _{HL}$=125 meV, $T_{e}$=$T_{h}$=2K, and few
magnitudes of magnetic field strength $h=-c\sqrt{2t}$: $c=0.4$
(curve 1), $c=0.8$ (curve 2), $c=1$ (curve 3) and $c=1.25$ (curve
4).}
\end{figure}

If HH splitting is not small as compared to temperature $T_{h}$ in some
range of angles $\varphi $, the Eq.(\ref{f17}) shows qualitatively different
character of OPA. Namely, the higher harmonics with large amplitudes can
appear. This can be regarded as a manifestation of higher powers in
expansion of $\tanh \left( \omega _{h}/2T_{h}\right) $. The Fig.2 reports
some calculated curves of OPA and corresponding effective $g$-factor
anisotropy demonstrating a new net effect without the influence of cubic
anisotropy of $V_{q}$.

Very interesting experimental data had been obtained in Ref.%
\cite{KusrPRL}, where second and visible fourth harmonics of the
OPA were detected in PL of 20 $\mathring{A}$ CdTe QW with semimagnetic $%
Cd_{1-x}Mn_{x}Te$ barriers. Exchange interaction with magnetic ions in the
barriers and interfaces enlarges HH splitting that makes possible to reach a
significant magnitude of HH and electron spin polarization at liquid helium
temperature.\cite{DMSQW} Quantitative analysis carried out in terms of Eqs (%
\ref{f17}) and (\ref{f16}) shows that parameters $h=-0.056$, $\omega
_{e}=0.014\Delta _{HL}$, $T_{e}=T_{h}=\Delta _{HL}/720$, $t=0.001$ and $%
q_{1}=-0.0006$ describe nicely the OPA experimental data of Ref.%
\cite{KusrPRL} (see Fig.3) assuming that $\Delta _{HL}=125$ meV
and electron and hole spin temperatures are equal to lattice
temperature $T=2K$.
These calculations explain also the effect of great OPA amplification from $%
0.1\%$ (that could not be detected in Ref.\cite{KusrPRL} because
of experimental errors) to $15\%$ \ under the magnetic field
action.

\begin{figure}[th]
\centering{\
\includegraphics[width=7cm]{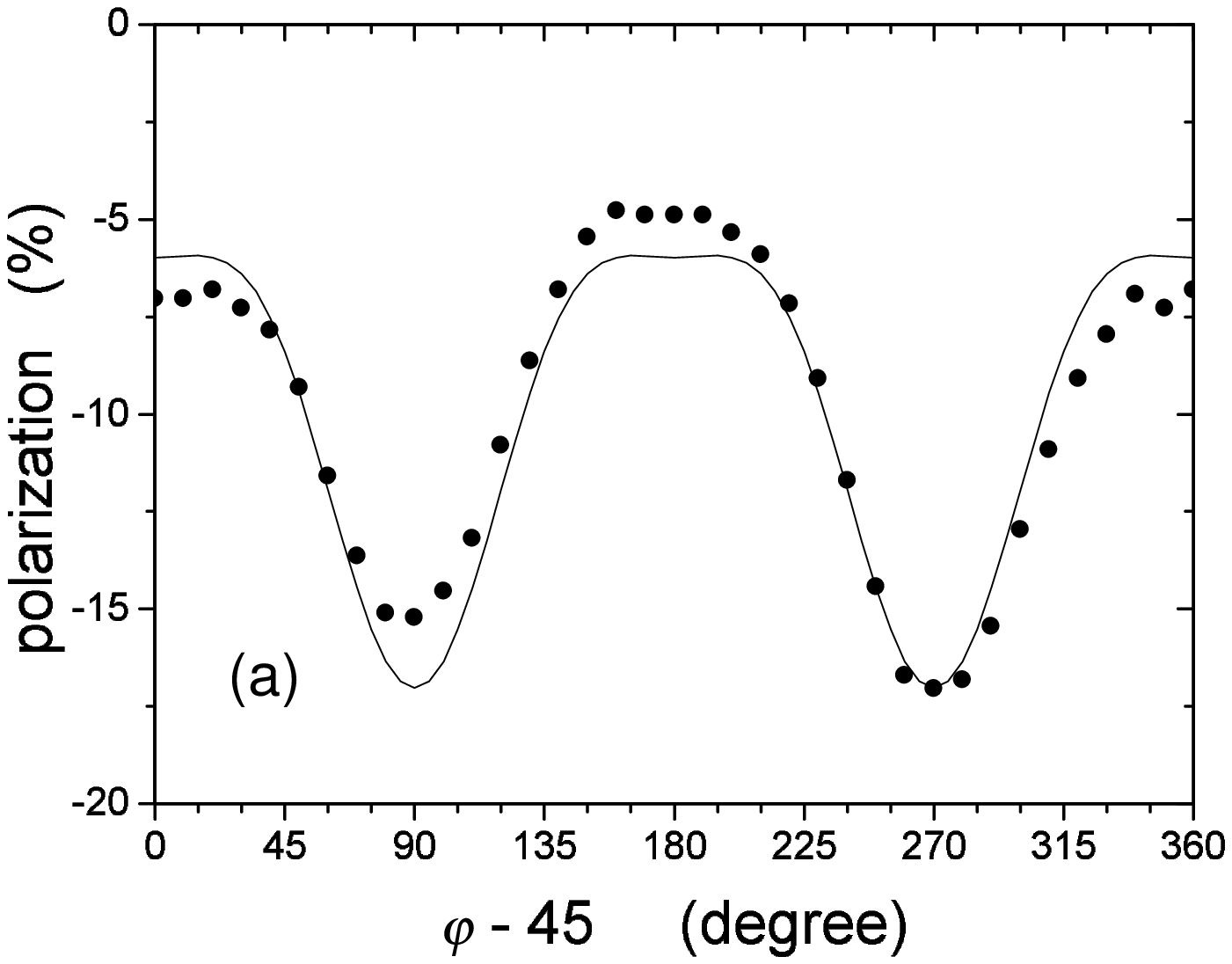}}
\vspace*{15mm}
\centering{\
\includegraphics[width=7cm]{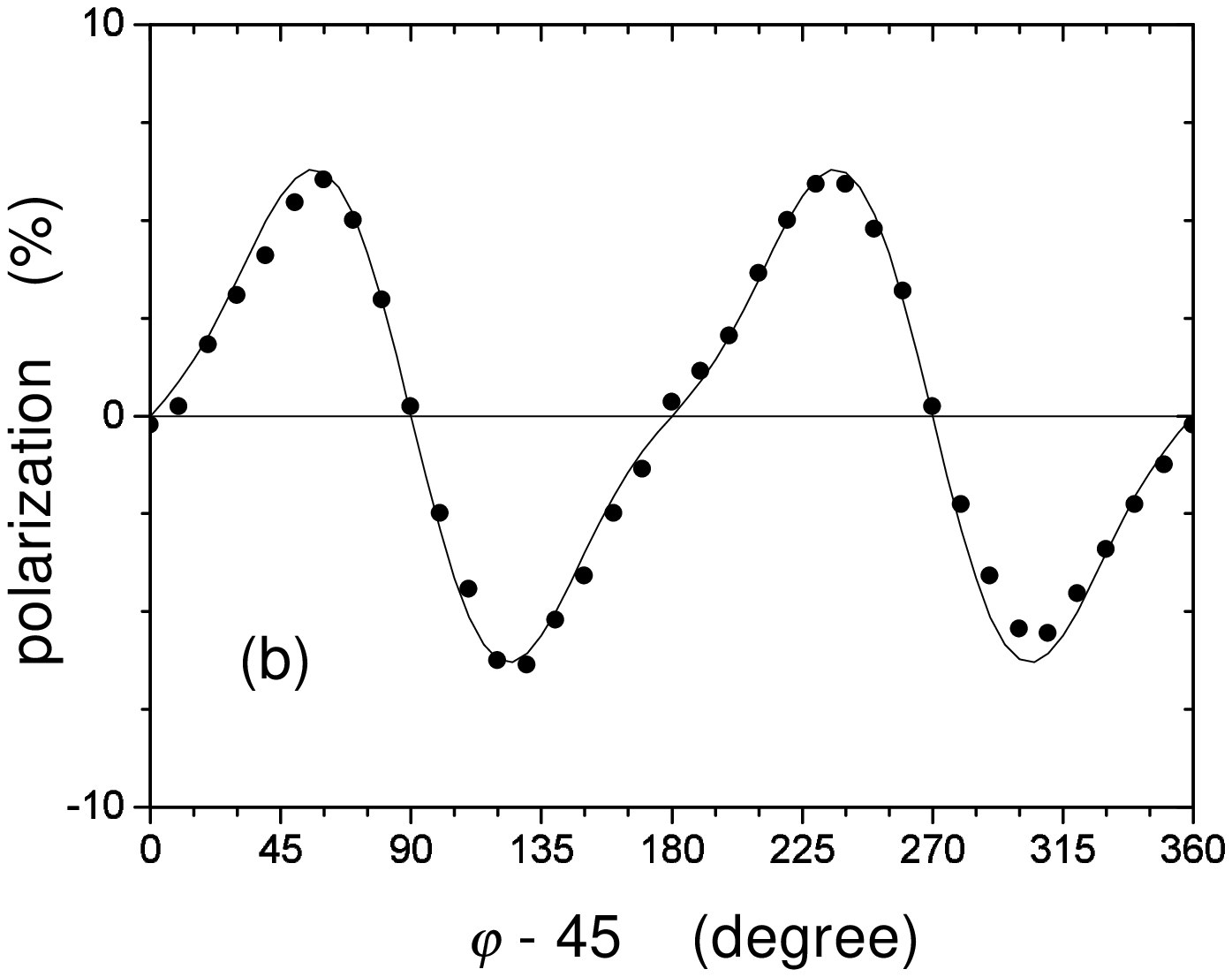}}
\caption{Comparison of the OPA calculated in terms of Eq.
(\ref{f17}) (solid
lines) with experimental data (points) of Kusrayev et al (Ref. %
\cite{KusrPRL}, Fig.3b therein) recorded in the parallel ($\alpha
=0^{\circ }$) to the magnetic field polarization plane (a) and plane with $%
\alpha =45^{\circ }$ rotated relative to $\alpha =0^{\circ }$ by
$45^{\circ } $ (b). Fitting parameters see in the text.}
\end{figure}

Successful description of experiment raises the question about the
relationship of above obtained parameters with those for QWs with other
widths. There are few mechanisms of QW width $L_{w}$ influence on OPA. (i)
Effective fields $G_{e}$ and $G_{h}$ originated from carriers exchange
interaction with magnetic ions are proportional to overlap of electron and
hole $\psi $ functions densities with semimagnetic barriers and interfaces.
One can expect significant decreasing of this exchange field in very wide
non-magnetic wells. On the other hand, the strengths of exchange fields can
have non-monotonic dependencies on QW width due to effect of giant
paramagnetic enhancement.\cite{DMSQW,Siviniant} However, we can expect a
reduction of a magnetic field influence on the OPA with QW broadening,
especially for $V_{Z}^{(3)}$ contribution (\ref{f9}), which is proportional
to $G_{h}^{3}G_{e}$. (ii) The HH-LH splitting depends primarily on $L_{w}$.
In the case of high QW barriers, one can expect $\Delta _{HL}\propto
1/L_{w}^{2}$ that increases the roles of terms $V_{Z}^{(3)}$, $V_{ht}^{(2)}$
and $V_{h\varkappa }^{(2)}$ with QW broadening but does not influence the
term $V_{q}^{(1)}$. This can be principal mechanism decreasing the role of
fourth harmonic with respect to second harmonic of OPA in Ref.%
\cite{KusrPRL}. (iii) Narrow QWs are most favorable for carrier
localization on the fluctuations of random potential as well as
for polaron formation. This increases the role of random potential
(\ref{r1}) that can suppress the OPA, as noted in Section III.

Aforementioned analysis explains qualitatively why fourth harmonic
of OPA is evident for most narrow QW in the experiment
Ref.\cite{KusrPRL}. On
the other hand, it shows that situation with dependence of the OPA on the $%
L_{w}$ may be very complicated. Most adequate approach to this
problem seems to consist in independent determination of $\Delta
_{HL}$ as well as $G_{e}$ and $G_{h}$ from magnetooptical
measurements. Then, microscopic constants in Hamiltonian
(\ref{f14}) should be found from comparison of general expressions
(\ref{f17}) with experimentally observed OPA. The cited data of
Ref. \cite{KusrPRL} do not allow to perform this program due to
the loss of $\Delta _{HL}$, $G_{e}$ and $G_{h}$ data.

\section{Conclusion}

We have developed a microscopic theory of OPA in QWs subjected to the
in-plane magnetic field. Two types of optical polarization contributions
should be distinguished. First is due to admixture of LH to HH states. This
effect is small as perturbation theory predicts. A HH splitting in a
magnetic field determines other type of polarization mechanism owing to
phase correlations of electron and hole $\psi $ functions. This effect can
lead to almost 100\% polarization for suitably distinguishable four spectral
lines of electron-HH optical transitions in spite of relatively small
interactions ($\ll \Delta _{HL}$) responsible for HH splitting. Besides, we
have considered spectral properties of OPA. These polarization peculiarities
turns out sensitive to PL (absorption, reflectivity, etc) lineshape in the
case of relatively small Zeeman splitting.

Theory considers Zeeman interaction, non Zeeman HH splitting and
$C_{2v}$ potentials as sources of different OPA. Their joint
manifestation reveals peculiar OPA behavior due to interference
effects. A random potential, localizing HHs should be considered
separately as a depolarization factor of the PL. We predicted some
new effects, (i) the anisotropy of HH splitting (or $g$-factor)
due to interference of different HH potentials, (ii) manifestation
of fourth and higher harmonic in OPA caused by only $C_{2v}$
potential (Fig.2), (iii) polarization suppression at the
conditions of crossing (anti-crossing) of HH levels, (iv)
depolarization effect of random potential influence. Our theory
gives full qualitative description for some important experimental
details of OPA found earlier.

The work was supported, in part, by the INTAS 99-015 grant and by
grant of State Fundamental Research Foundation of Ukraine
02.07/0125.


\begin{thebibliography}{99}
\bibitem{comm2} In the case of $B=0$ we assume that infinitesimal ${\vec{B}}$
determines only a plane that forms an angle $\varphi $ with $\overrightarrow{%
OX}$ direction.
\bibitem{comm3} In Ref.\cite{KusrPRL} the principal axes $x''$ and $y''$
of $C_{2v}$ interaction rotated relative $%
x'$ and $y'$ by $45^{\circ}$ about the $z$ axis are used:
$g_{x''x''}^{\perp }=-g_{y''y''}^{\perp }$, $%
g_{x''y''}^{\perp }=g_{y''x''}^{\perp }=0$.
\end{thebibliography}
\end{document}